\documentclass[12pt]{article}
\usepackage{mathtext}
\usepackage[cp1251]{inputenc}
\usepackage{amsmath}
\usepackage{bm}
\usepackage[english,ukrainian]{babel}
\usepackage{color}
\usepackage{cite}
\begin{document}
\begin{center}
\begin{large}{\bf  Particle in uniform field in noncommutative space with preserved time reversal  and rotational symmetries}\end{large}
\end{center}

\centerline {Kh. P. Gnatenko\footnote{E-Mail address: khrystyna.gnatenko@gmail.com}, Kh. I. Stakhur,  A. V. Kryzhova}
\medskip
\centerline {\small \it Department for Theoretical Physics, Ivan Franko National University of Lviv}
\centerline {\small \it 12 Drahomanov Str., Lviv, 79005, Ukraine}

\medskip
\begin{center}
{\bf Abstract}
\end{center}
Quantized space described by time reversal invariant and rotationally invariant noncommutative algebra of canonical type is studied. A particle in uniform field is considered. We find exactly  the energy of a particle in uniform field  in the quantized space and its wavefunctions. It is shown that the motion of the particle  in the field direction in time reversal invariant and rotationally invariant noncommutative space is the same as in the ordinary space (space with the ordinary commutation relations for operators of coordinates and operators of momenta). Noncommutativity of coordinates has influence only on the motion of the particle in the directions perpendicular to the field direction. Namely, space quantization has effect on the mass of the particle.

  {\bf Key words:} quantized space, noncommutative coordinates, time reversal symmetry, rotational symmetry, particle in uniform field

  PACS number(s): 02.40.Gh, 03.65.-w

\section{Introduction}
According to the  String theory and Quantum gravity a minimal length of the order of the Planck length exists \cite{Witten,Doplicher}. A space with minimal length can be described on the basis of idea of deformation of the commutation relations for  operators of coordinates and operators of momenta.

Many different deformed algebras leading to the minimal length were proposed. One can distinguish three types of the algebras: noncommutative algebras of canonical type (see, for instance, \cite{Gamboa2001,Nair2001,Chaichian2001,Biswas2020}), noncommutative algebras of Lie type (see, for instance, \cite{Lukierski2006,Amelino-Camelia2002,Daszkiewicz2008}), nonlinear deformed algebras (see, for instance, \cite{Chang2002,Benczik2005,Tkachuk2010}). Noncommutative algebras of canonical type are the most simple algebras which describe space quantization on the  Planck scale.
Algebra with noncommutativity of coordinates of canonical type is characterized by the following relations
  \begin{eqnarray}
[X_{i},X_{j}]=i\hbar\theta_{ij},\label{form101}\\{}
[X_{i},P_{j}]=i\hbar\delta_{ij},\label{form1001}\\{}
[P_{i},P_{j}]=0.\label{form10001}{}
\end{eqnarray}
Here $\theta_{ij}$ are called parameters of noncommutativity which  are elements of constant  matrixes. Properties of physical systems in the frame of noncommutative algebra of canonical type have been widely studied (see, for instance, \cite{Gamboa2001,Nair2001,Chaichian2001,Biswas2020,Schneider2020, Bertolami2005, Bertolami2011, Romero, Mirza,BastosPLA,Castello2010}, and references therein). Among the problems noncommutative gravitational quantum well has been examined  \cite{Bertolami2005,BastosPLA,Castello2010}.
It is worth stressing that in noncommutative space characterized by commutation relations (\ref{form101})-(\ref{form10001}) the rotational and time reversal symmetries are not preserved \cite{Chaichian2001,Geloun,Scholtz,GnatenkoPRA}. In \cite{Gnatenko14} noncommutative algebra which is rotationally invariant and equivalent to nocommutative algebra of canonical type was proposed. In \cite{GnatenkoMPLA16} effect of noncommutativity on the mass of a particle in uniform field was found in the frame of the rotationally invariant noncommutative algebra of canonical type \cite{Gnatenko14}.

In the present paper we study a particle in uniform filed in the frame of rotationally invariant and time reversal invariant noncommutative algebra of canonical type proposed in \cite{GnatenkoPRA}. The total Hamiltonian is constructed and analyzed. We find exactly energy and wave functions  of a particle in uniform field in noncommutative space with preserved rotational and time reversal symmetries.

The paper is organized as follows. In Section 2 time reversal and rotationally invariant algebra with noncommutativity of coordinates is presented. Section 3 is devoted to studies of a particle in uniform field in rotationally invariant and time reversal invariant noncommutative space. Conclusions are presented in Section 4.

\section{Time reversal and rotationally invariant algebra with canonical noncommutativity of coordinates}
In \cite{GnatenkoPRA} for preserving rotational and time reversal symmetries in noncommutative space the authors considered the idea to generalize parameters of noncommutativity, defining tensor of coordinate noncommutativity as
 \begin{eqnarray}
\theta_{ij}=\frac{c_{\theta}}{\hbar}\sum_k\varepsilon_{ijk}{p}^a_{k},\label{t1}
\end{eqnarray}
here ${p}^{a}_i$ are additional momenta, governed by a rotationally invariant system, $c_{\theta}$ is a constant. For the reason of simplicity in \cite{GnatenkoPRA} the additional momenta  were assumed to be governed by harmonic oscillator
\begin{eqnarray}
H^a_{osc}=\frac{({\bf p}^{a})^{2}}{2m_{osc}}+\frac{m_{osc}\omega^2_{osc}{\bf a}^{2}}{2}.\label{harn}
\end{eqnarray}
Here $a_i$ are additional coordinates conjugated to momenta $p_i^a$. The following relations are satisfied
\begin{eqnarray}{}
[{a}_{i},{p}^{a}_{j}]=i\hbar\delta_{ij},\\{}
[{a}_{i},{a}_{j}]=[{p}^{a}_{i},{p}^{a}_{j}]=0.
\end{eqnarray}

The oscillator length is considered to be equal to the Planck length $\sqrt{\hbar}/{\sqrt{m_{osc}\omega_{osc}}}=l_P$ and the frequency of the oscillator is assumed to be very large \cite{GnatenkoPRA}. So, the harmonic oscillator (\ref{harn}) remains in the ground state.

The time reversal and rotationally invariant algebra with canonical noncommutativity of coordinates reads
\begin{eqnarray}
[X_{i},X_{j}]=ic_{\theta}\sum_k\varepsilon_{ijk}{p}^a_{k},\label{for101}\\{}
[X_{i},P_{j}]=i\hbar\delta_{ij},\ \
[P_{i},P_{j}]=0, \label{for10001}\\{}
[{p}^{a}_{i},X_{j}]=[{p}^{a}_{i},P_{j}]=0.\label{f11}{}
\end{eqnarray}

It is convenient to represent coordinates and momenta which satisfy  (\ref{for101}), (\ref{for10001}) by coordinates and momenta $x_i$, $p_i$ satisfying the ordinary commutation relations
\begin{eqnarray}
[x_i,x_j]=[p_i,p_j]=0,\label{r}\\{}
[x_i,p_j]=i\hbar\delta_{ij}.\label{rr}
\end{eqnarray}
 The representation is
 \begin{eqnarray}
X_{i}=x_{i}+\frac{1}{2}[{\bm \theta}\times{\bf p}]_i, \ \
P_{i}=p_{i},\label{repp0}
\end{eqnarray}
here  ${\bm \theta}=(\theta_1,\theta_2,\theta_3)$,
 \begin{eqnarray}
\theta_i=\frac{c_{\theta}{p}_i^a}{\hbar}.\label{th}
 \end{eqnarray}

From relation  (\ref{f11}) follows that  algebra (\ref{for101}), (\ref{for10001}) is equivalent to algebra with canonical noncommutativity of coordinates.
 Relations (\ref{for101}), (\ref{for10001}) are invariant under the time reversal transformation which include complex conjugation. Also, after this transformation the coordinates and the momenta change as $X_{i}\rightarrow X_{i}$, $P_{i}\rightarrow -P_{i}$, $p^a_{i}\rightarrow -p^a_{i}$.  Algebra (\ref{for101}), (\ref{for10001}) is time reversal invariant \cite{GnatenkoPRA}.
After rotation the coordinates and  momenta change as $X_{i}^{\prime}=U(\varphi)X_{i}U^{+}(\varphi)$, $P_{i}^{\prime}=U(\varphi)P_{i}U^{+}(\varphi)$ $p^{a\prime}_{i}=U(\varphi)p^a_{i}U^{+}(\varphi)$, $U(\varphi)=\exp(i\varphi({\bf n}\cdot{\bf L^t})/\hbar)$ with ${\bf L^t}=[{\bf x}\times{\bf p}]+[{\bf{a}}\times{\bf {p}}^{a}]$.
Commutation relations (\ref{for101}), (\ref{for10001}) are invariant under rotation, algebra is rotationally invariant \cite{GnatenkoPRA}.

In the next section we study a particle in uniform field in the frame of rotationally invariant and time reversal  invariant noncommutative algebra (\ref{for101}), (\ref{for10001}).

\section{Energy of a particle in uniform field in rotationally invariant and time reversal invariant noncommutative space}
Let us study a particle with mass $m$ in uniform field
with the following Hamiltonian
 \begin{eqnarray}
 H_{p}=\frac{P^{2}}{2m}-\alpha X_{3}.
 \label{form888}
  \end{eqnarray}
   In (\ref{form888}) coordinate and momenta satisfy relations (\ref{for101}), (\ref{for10001}). Without loss of generality, for convenience we consider the field pointed in the $X_{3}$ direction (in (\ref{form888}) $\alpha$ characterize the force acting on the particle). Because algebra (\ref{for101}), (\ref{for10001}) is rotationally invariant the results  of this section can be generalized to the case of arbitrary direction of the field.

To construct time reversal invariant and rotationally invariant noncommutative algebra (\ref{for101}), (\ref{for10001}) additional momenta $p_i^a$ were involved, therefore to study a particle in uniform filed in the space (\ref{for101}), (\ref{for10001}) one should write the following Hamiltonian
\begin{eqnarray}
H=\frac{P^{2}}{2m}-\alpha X_{3}+\frac{(p^{a})^{2}}{2m_{osc}}+\frac{m_{osc}\omega_{osc}^{2}a^{2}}{2},\label{form1333}
\end{eqnarray}
the last two terms in which correspond to harmonic oscillator (\ref{harn}). Then to find influence of space quantization  on the energy of a particle in uniform field it is convenient to use representation (\ref{repp0}) and rewrite Hamiltonian (\ref{form1333}) as
follows
\begin{eqnarray}
 H=\frac{p^{2}}{2m}-\alpha x_{3}-\frac{1}{2}[{\bm \theta}\times{\bf p}]_3
  +\frac{(p^{a})^{2}}{2m_{osc}}+\frac{m_{osc}\omega_{osc}^{2}a^{2}}{2}=\nonumber\\=\frac{p^{2}}{2m}-\alpha x_{3}-\frac{\alpha c_{\theta}}{2\hbar}(p_1^ap_2-p_2^ap_1)
  +\frac{(p^{a})^{2}}{2m_{osc}}+\frac{m_{osc}\omega_{osc}^{2}a^{2}}{2}\label{form600}
\end{eqnarray}
Here we take into account (\ref{th}).

To find exact expression  for the energy of a particle in uniform field in space described by commutation relations (\ref{for101}), (\ref{for10001}), we rewrite Hamiltonian (\ref{form600}) as
\begin{eqnarray}
 H=\left(1-\frac{\alpha^{2}c_{\theta}^{2}m m_{osc}}{4\hbar^{2}}\right)\frac{p_{1}^{2}}{2m}+\left(1-\frac{\alpha^{2}c_{\theta}^{2}m m_{osc}}{4\hbar^{2}}\right)\frac{p_{2}^{2}}{2m}+ \frac{p_{3}^{2}}{2m}-\alpha x_{3}+\nonumber\\+\frac{1}{2m_{osc}}\left(p^{a}_1-\frac{\alpha c_{\theta}m_{osc}}{2\hbar}p_2\right)^2+\frac{1}{2m_{osc}}\left(p^{a}_2+\frac{\alpha c_{\theta}m_{osc}}{2\hbar}p_1\right)^2+\nonumber\\+\frac{(p_3^a)^2}{2m_{osc}}+
 +\frac{m_{osc}\omega_{osc}^{2}a_1^{2}}{2}+ \frac{m_{osc}\omega_{osc}^{2}a_2^{2}}{2}+ \frac{m_{osc}\omega_{osc}^{2}a_3^{2}}{2}.\label{form60110}
\end{eqnarray}

Note that operators
\begin{eqnarray}
\tilde{H}_p=\left(1-\frac{\alpha^{2} c_{\theta}^{2}m }{4\hbar \omega_{osc} l^2_P}\right)\frac{p_{1}^{2}}{2m}+\left(1-\frac{\alpha^{2} c_{\theta}^{2}m }{4\hbar \omega_{osc} l^2_P}\right)\frac{p_{2}^{2}}{2m}+ \frac{p_{3}^{2}}{2m}-\alpha x_{3},\label{th1}
\end{eqnarray}
and
\begin{eqnarray}
\tilde{H}_{osc}=\frac{1}{2m_{osc}}\left(p^{a}_1-\frac{\alpha c_{\theta}}{2\omega_{osc}l^2_P}p_2\right)^2+\frac{1}{2m_{osc}}\left(p^{a}_2+\frac{\alpha c_{\theta}}{2\omega_{osc}l^2_P}p_1\right)^2+\nonumber\\+\frac{(p_3^a)^2}{2m_{osc}}+
 \frac{m_{osc}\omega_{osc}^{2}a_1^{2}}{2}+ \frac{m_{osc}\omega_{osc}^{2}a_2^{2}}{2}+ \frac{m_{osc}\omega_{osc}^{2}a_3^{2}}{2},\label{th2}
\end{eqnarray}
in (\ref{form60110}) commute
\begin{eqnarray}
[\tilde{H}_p, \tilde{H}_{osc}]=0.\label{com1}
\end{eqnarray}
 Writing (\ref{th1}), (\ref{th2}) we take into account that
\begin{eqnarray}
\sqrt{\frac{\hbar}{m_{osc}\omega_{osc}}}=l_P,
\end{eqnarray}
 as was assumed in the paper  \cite{GnatenkoPRA}, where the noncommutative algebra invariant upon time reversal and rotationally invariant was constructed.

 Operator $\tilde{H}_p$ can be rewritten as
\begin{eqnarray}
\tilde{H}_p=\tilde{H}_{1}+\tilde{H}_{2}+\tilde{H}_{3},
\end{eqnarray}
with
\begin{eqnarray}
\tilde{H}_{1}=\frac{p_{1}^{2}}{2m_{eff}},\label{uf1}\\
\tilde{H}_{2}= \frac{p_{2}^{2}}{2m_{eff}},\label{uf2} \\
\tilde{H}_{3}=\frac{p_{3}^{2}}{2m}-\alpha x_{3},\label{uf3}\\{}
[\tilde{H}_{1},\tilde{H}_{2}]=[\tilde{H}_{2},\tilde{H}_{3}]=[\tilde{H}_{1},\tilde{H}_{3}]=0.\label{comm}
\end{eqnarray}
The effective mass reads
\begin{eqnarray}
m_{eff}=m\left(1-\frac{\alpha^{2} c_{\theta}^{2}m m_{osc}}{4\hbar^{2}}\right)^{-1}=m\left(1-\frac{\alpha^{2} c_{\theta}^{2}m }{4\hbar \omega_{osc} l^2_P}\right)^{-1}.\label{form1}
\end{eqnarray}

Note, that operator of coordinate $x_3$ and operator of momentum $p_3$ in $\tilde{H}_{3}$ satisfy the ordinary commutation relations (\ref{r}), (\ref{rr}).  So,  $\tilde{H}_{3}$ corresponds to the Hamiltonian of a particle in uniform filed in the ordinary space  (in a space in which the operators of coordinates and operators of momenta satisfy the ordinary commutation relations).

Introducing
\begin{eqnarray}
\tilde{p}^a_{1}=p^{a}_1-\frac{\alpha c_{\theta}}{2\omega_{osc}l^2_P}p_2,\\
\tilde{p}^a_{2}=p^{a}_2+\frac{\alpha c_{\theta}}{2\omega_{osc}l^2_P}p_1,\\
\tilde{p}^a_{3}=p^{a}_3,
\end{eqnarray}
one can write (\ref{th2}) as
\begin{eqnarray}
\tilde{H}_{osc}=\frac{(\tilde p^{a})^{2}}{2m_{osc}}+\frac{m_{osc}\omega_{osc}^{2}a^{2}}{2}.\label{form2}
\end{eqnarray}

Operators $a_i$ and $\tilde{p}^a_{i}$ satisfy the ordinary commutation relations
\begin{eqnarray}
[a_{i},a_{j}]=[\tilde{p}^a_{i},\tilde{p}^a_{j}]=0,\\{}
[a_i,\tilde{p}^a_{j}]=i\hbar\delta_{ij},
\end{eqnarray}
Therefore operator (\ref{form2}) corresponds to Hamiltonian of three-dimensional harmonic oscillator with mass $m_{osc}$ and frequency $\omega_{osc}$ in the ordinary space. Spectrum of the oscillator in the ordinary space is well known. Let us recall that the frequency $\omega_{osc}$ is very large  \cite{GnatenkoPRA}, and oscillator putted in the ground state remains in it.  So, the oscillator energy is $3\hbar\omega_{osc}/{2}$.

Operators $\tilde{H}_{1}$, $\tilde{H}_{2}$, $\tilde{H}_{3}$, $\tilde{H}_{osc}$ commute with each other (see (\ref{com1}), (\ref{comm})). So, the spectrum of a particle in uniform filed in rotationally invariant and time reversal invariant noncommutative space reads
\begin{eqnarray}
E=\frac{\hbar^2k_1^2}{2m}\left(1-\frac{\alpha^{2} c_{\theta}^{2}m }{4\hbar \omega_{osc} l^2_P}\right)+\frac{\hbar^2k_2^2}{2m}\left(1-\frac{\alpha^{2} c_{\theta}^{2}m }{4\hbar \omega_{osc} l^2_P}\right)+E_3+\frac{3}{2}\hbar\omega_{osc}.\label{form20000}
\end{eqnarray}
Note that the motion of a particle in the directions perpendicular to the field direction is free.
In (\ref{form20000}) $k_1$, $k_2$ denote components of the wave vector which correspond to this motion,  $E_3$  denotes continious eigenvalues of $\tilde{H}_{3}$.
 The last term in (\ref{form20000}) corresponds to the ground state energy of the harmonic oscillator (\ref{form2}).

Let us also write the eigenfunctions of the total Hamiltonian (\ref{form60110}).
Because relations (\ref{com1}), (\ref{comm}) are satisfied, we can write
 \begin{eqnarray}
\psi({\bf x},{\bf a})=\tilde{\psi}_1(x_1)\tilde{\psi}_2(x_2)\tilde{\psi}_3(x_3)\tilde{\psi}({\bf a})\label{psi}
\end{eqnarray}
where $\tilde{\psi}_i(x_i)$ are eigenfunctions of $\tilde{H}_i$ given by (\ref{uf1})-(\ref{uf3}). Note that $\psi^{(3)}(x_3)$  is eigenfunction of a particle in uniform field in the ordinary space, which is well known (see, for instance, \cite{1}). It reads
\begin{eqnarray}
\psi^{(3)}(x_3)=\left(\frac{4m^2}{\pi^3\alpha\hbar^{4}}\right)^{\frac{1}{6}}\Phi\left(\left(\frac{2m\alpha}{\hbar^2}\right)^{\frac{1}{3}}\left(-x_3-\frac{E_3}{\alpha}\right)\right),
\end{eqnarray}
here $\Phi$ is the Airy function
\begin{eqnarray}
\Phi(x)=\frac{1}{\sqrt{\pi}}\int_0^{\infty}\cos\left(\frac{t^3}{3}+tx\right)dt.
\end{eqnarray}
 Functions $\tilde{\psi}({\bf a})$ in (\ref{psi}) are eigenfunctions of
\begin{eqnarray}
{H}^{\prime}_{osc}=\frac{1}{2m_{osc}}\left(p^{a}_1-\frac{\alpha c_{\theta}\hbar k_2}{2\omega_{osc}l^2_P}\right)^2+\frac{1}{2m_{osc}}\left(p^{a}_2+\frac{\alpha c_{\theta}\hbar k_1}{2\omega_{osc}l^2_P}\right)^2+\nonumber\\+\frac{(p_3^a)^2}{2m_{osc}}+
 \frac{m_{osc}\omega_{osc}^{2}a_1^{2}}{2}+\frac{m_{osc}\omega_{osc}^{2}a_2^{2}}{2}+ \frac{m_{osc}\omega_{osc}^{2}a_3^{2}}{2}.\label{harm3}
\end{eqnarray}
Hamiltonian (\ref{harm3}) obtained replacing   $p_1$ and $p_2$ by $\hbar k_1$, $\hbar k_2$, respectively,  in (\ref{th2}).
The eigenfunction of (\ref{harm3}) corresponding to the ground state reads
\begin{eqnarray}
\tilde{\psi}({\bf a})=\frac{1}{\pi^{\frac{3}{4}} l_P^{\frac{3}{2}}}e^{-\frac{a^2}{2l^2_P}-i\beta(k_1a_2-k_2a_1)}.
\end{eqnarray}
Here for convenience we use the following notation
\begin{eqnarray}
\beta=\frac{\alpha c_{\theta}}{2\omega_{osc}l^2_P}.
\end{eqnarray}

So, we can write the eigenfunctions of the total Hamiltonian (\ref{form60110}). They  read
\begin{eqnarray}
\psi({\bf x},{\bf a})=C e^{ik_1x_1}e^{ik_2x_2}\Phi\left(\left(\frac{2m\alpha}{\hbar^2}\right)^{\frac{1}{3}}\left(-x_3-\frac{E_3}{\alpha}\right)\right) e^{-\frac{a^2}{2l^2_P}-i\beta(k_1a_2-k_2a_1)},\label{eigf}
\end{eqnarray}
where $C$ is the normalization constant.

Let us analyze the obtained results. It is important to note that features of space structure on the Planck scale have effect only on the motion of a particle in the directions perpendicular to the direction of the field.  The first two terms in  (\ref{form20000}) can be rewritten by effective mass (\ref{form1}).
 So, space quantization has  effect  on the mass of the particle in uniform field in rotationally invariant and time reversal invariant space with noncommutativity of coordinates.

\subsection*{Conclusion}

In the paper we have considered algebra with noncommutativity of coordinates which is rotationally and time reversal invariant (\ref{for101}), (\ref{for10001}). This algebra describes space quantization at the Planck scale. Influence of space quantization on the motion of a particle in uniform field has been studied. Taking into account that the rotationally invariant and time reversal invariant noncommutative algebra contains additional momenta,   we have constructed and examined total  Hamiltonian of a particle in uniform field in time reversal invariant and rotationally invariant noncommutative space (\ref{form1333}). Energy and wave functions of the particle have been found exactly (\ref{form20000}), (\ref{eigf}).  We obtain that the motion of a particle in the field direction in rotationally and time-reversal invariant noncommutative space is the same as in the space with ordinary commutation relations for operators of coordinates and operators of momenta.  Features of space structure described by noncommutative algebra (\ref{for101}), (\ref{for10001}) have effect on the motion of a particle in the directions perpendicular to the field direction.  Similarly as in the ordinary space the motion of a particle in these directions is free. The noncommutativity has only effect on the particle mass.

\subsection*{Acknowledgement}
The authors thank Prof. Tkachuk V. M. for his advices and support during research studies.  This work was partly supported by the Project $\Phi\Phi$-11Hp (No. 0121U100058) from the Ministry of Education and Science of Ukraine.

\newpage
\begin{center}
\begin{large}{\bf Частинка в однорідному полі у некомутативному просторі зі збереженою симетрією відносно інверсії часу та сферичною симетрією}
\end{large}
\end{center}

\centerline { Х. П. Гнатенко, Х. І. Стахур, А. В. Крижова}
\medskip

\centerline {\small  \it Кафедра теоретичної фізики,}
\centerline {\small \it Львівський національний університет імені Івана Франка,}
\centerline {\small \it вул. Драгоманова, 12, 79005 Львів, Україна}

\medskip
\centerline {{\bf Анотація}}
Вивчається вплив квантованості простору  на  рух частинки у однорідному полі. Для  опису особливостей структури простору на планківських масштабах та врахування квантованості простору розглядається ідея деформації звичних комутаційних співвідношень для операторів координат. А саме, припускається, що комутатор координат не дорівнює нулеві. У літературі відомо багато різних типів деформацій звичних комутаційних співвідношень для операторів координат та імпульсів. Найпростішою та найбільш вивченою є алгебра з некомутативнісню координат канонічного типу. Така алгебра описує квантований простір, але зумовлює порушення симетрії відносно інверсії часу, порушення сферичної симетрії. У статті вивчається інваріантна відносно інверсії часу та сферично-симетрична алгебра з некомутативністю координат  канонічного типу, яка описує простір з мінімальною довжиною (квантований простір) та була запропонована у роботі [Kh. P. Gnatenko, M. I. Samar, V. M. Tkachuk, Phys. Rev. A 99, 012114 (2019)]. Алгебра характеризується тензором некомутативності, побудованим за допомогою додаткових імпульсів. Останні відповідають сферично-симетричній системі. Розглядається випадок, коли ця система – це гармонічний осцилятор з довжиною, яка дорівнює довжині Планка, та великою частотою. Досліджується частинка в однорідному полі у квантованому просторі зі збереженою симетрією відносно інверсії часу та сферичною симетрією.  Ми побудували та проаналізували повний гамільтоніан частинки в однорідному полі. Використавши метод представлення некомутативних координат через координати та імпульси, які задовольняють звичні комутаційні співвідношення, ми знайшли  точний вираз для енергії та хвильових функцій частинки у однорідному полі. У статті показано, що рух частинки  у напрямку однорідного поля у сферично-симетричному та інваріантному відносно інверсії часу просторі з некомутативністю координат   є таким самим, як у звичному просторі. Некомутативність координат впливає тільки на рух частинки в напрямках перпендикулярних до поля. А саме, квантованість простору впливає на масу частинки.

\end{document}